\documentclass[sigconf]{acmart}
\acmConference[ISSTA 2023]{ACM SIGSOFT International Symposium on Software Testing and Analysis}{17-21 July, 2023}{Seattle, USA}

\AtBeginDocument{%
  }

\setcopyright{none}

\usepackage{algorithmic}
\usepackage[ruled,linesnumbered,vlined]{algorithm2e}
\usepackage{setspace}
\usepackage{listings}
\usepackage{multicol}
\usepackage{multirow}
\usepackage{makecell}
\usepackage{amsthm}

\usepackage[font=footnotesize,labelfont=bf]{subcaption}
\newcommand{\parabf}[1]{\noindent\textbf{#1}}

\newcommand{\mypara}[1]{\vspace{.03in}\noindent \textbf{#1}}
\newcommand{\CodeIn}[1]{{\small \texttt{#1}}}
\newcommand{\Comment}[1]{}

\newcommand{\tech}{\textsc{TitanFuzz}\xspace} %

\newcommand{\cradle}{CRADLE\xspace}
\newcommand{\lemon}{LEMON\xspace}

\newcommand{\audee}{AUDEE\xspace}
\newcommand{\muffin}{Muffin\xspace}
\newcommand{\freefuzz}{FreeFuzz\xspace}
\newcommand{\docter}{DocTer\xspace}
\newcommand{\deeprel}{DeepREL\xspace}

\newcommand{\nnsmith}{NNSmith\xspace}
\newcommand{\nablafuzz}{$\nabla$Fuzz\xspace}

\newcommand{\tf}{TensorFlow\xspace}
\newcommand{\pt}{PyTorch\xspace}
\newcommand{\python}{Python\xspace}

\newcommand{\plm}{LLM\xspace} %
\newcommand{\plmfull}{Large Pre-trained Language Model\xspace}
\newcommand{\nlp}{NLP\xspace}
\newcommand{\nlpfull}{Natural Language Processing\xspace}

\newcommand{\codex}{Codex\xspace} 
\newcommand{\incoder}{\textsc{InCoder}\xspace}

\newcommand{\generative}{generative\xspace}
\newcommand{\Generative}{Generative\xspace}
\newcommand{\infilling}{infilling\xspace}
\newcommand{\Infilling}{Infilling\xspace}

\newcommand{\randombaseline}{Random\xspace}
\newcommand{\coveragebaseline}{Coverage\xspace}

\newcommand{\prefix}{prefix\xspace}
\newcommand{\suffix}{suffix\xspace}
\newcommand{\argument}{argument\xspace}
\newcommand{\keyword}{keyword-insertion\xspace}
\newcommand{\method}{method\xspace}

\newcommand{\argumentmut}{argument-replacement\xspace}

\newcommand{\prefixonlymut}{prefix-only\xspace}
\newcommand{\suffixonlymut}{suffix-only\xspace}
\newcommand{\prefixargmut}{prefix-argument\xspace}
\newcommand{\suffixargmut}{suffix-argument\xspace}

\newcommand{\dagdepth}{D\xspace}
\newcommand{\apicall}{U\xspace}
\newcommand{\repeatedapicall}{R\xspace}

\newcommand{\numPtTotalAPI}{1593\xspace}
\newcommand{\numPtCoverAPI}{1329\xspace}
\newcommand{\numPtCoverAPIImprove}{24.09\%\xspace}

\newcommand{\numPtTotalBugs}{37\xspace}
\newcommand{\numPtConfirmedBugs}{31\xspace}
\newcommand{\numPtUnknownBugs}{23\xspace}

\newcommand{\numPtFixBugs}{6\xspace}
\newcommand{\numPtRejectBugs}{5\xspace}

\newcommand{\ptCoverage}{20.98\%\xspace}
\newcommand{\ptCoverageImprove}{50.84\%\xspace}

\newcommand{\numTFTotalAPI}{3316\xspace}
\newcommand{\numTFCoverAPI}{2215\xspace}
\newcommand{\numTFCoverAPIImprove}{91.11\%\xspace}

\newcommand{\numTFTotalBugs}{28\xspace}
\newcommand{\numTFConfirmedBugs}{22\xspace}
\newcommand{\numTFUnknownBugs}{18\xspace}

\newcommand{\numTFFixBugs}{2\xspace}
\newcommand{\numTFRejectBugs}{3\xspace}

\newcommand{\tfCoverage}{39.97\%\xspace}
\newcommand{\tfCoverageImprove}{30.38\%\xspace}

\newcommand{\numTotalBugs}{65\xspace}
\newcommand{\numConfirmedBugs}{53\xspace}
\newcommand{\numUnknownBugs}{41\xspace}

\newcommand{\numFixBugs}{8\xspace}
\newcommand{\numRejectBugs}{8\xspace}
\newcommand{\numConfirmedCodex}{10\space}
\newcommand{\numConfirmedBugsPrevWorkCanFind}{9\space}

\definecolor{mygreen}{rgb}{0,0.6,0}
\newcommand{\distance}{5pt}
\newtheoremstyle{exampstyle}
  {\topsep} %
  {\topsep} %
  {} %
  {} %
  {\bfseries} %
  {.} %
  {.5em} %
  {} %

\newtheorem{definition}{Definition}

\begin{document}

\title{Large Language Models are Zero-Shot Fuzzers: \\Fuzzing Deep-Learning Libraries via Large Language Models
}

\author{Yinlin Deng}
    \affiliation{\institution{University of Illinois Urbana-Champaign}\country{}}
    \email{yinlind2@illinois.edu}
\author{Chunqiu Steven Xia}
    \affiliation{\institution{University of Illinois Urbana-Champaign}\country{}}
    \email{chunqiu2@illinois.edu}
\author{Haoran Peng}
    \affiliation{\institution{University of Science and\\ Technology of China}\country{}}
    \email{hurrypeng@mail.ustc.edu.cn}
\author{Chenyuan Yang}
    \affiliation{\institution{University of Illinois Urbana-Champaign}\country{}}
    \email{cy54@illinois.edu}
\author{Lingming Zhang}
    \affiliation{\institution{University of Illinois Urbana-Champaign}\country{}}
    \email{lingming@illinois.edu}

\begin{abstract}

Deep Learning (DL) systems have received exponential growth in popularity and have become ubiquitous in our everyday life. Such systems are built on top of popular DL libraries, e.g., \tf and \pt which provide APIs as building blocks for DL systems. Detecting bugs in these DL libraries is critical for almost all downstream DL systems in ensuring effectiveness/safety for end users. Meanwhile, traditional fuzzing techniques can be hardly effective for such a challenging domain since the input DL programs need to satisfy both the input language (e.g., Python) syntax/semantics and the DL API input/shape constraints for tensor computations.

To address these limitations, we propose \tech{} – the first approach to directly leveraging Large Language Models (LLMs) to generate input programs for fuzzing DL libraries. LLMs are titanic models trained on billions of code snippets and can auto-regressively generate human-like code snippets. Our key insight is that modern LLMs can also include numerous code snippets invoking DL library APIs in their training corpora, and thus can implicitly learn both language syntax/semantics and intricate DL API constraints for valid DL program generation. More specifically, we use both generative and infilling LLMs (e.g., \codex /\incoder) to generate and mutate valid/diverse input DL programs for fuzzing. Our experimental results demonstrate that \tech{} can achieve \tfCoverageImprove{}/\ptCoverageImprove higher code coverage than state-of-the-art fuzzers on \tf{}/\pt. Furthermore, \tech is able to detect \numTotalBugs bugs, with \numUnknownBugs already confirmed as previously unknown bugs.

This paper demonstrates that modern titanic LLMs can be leveraged to \textit{directly} perform both generation-based and mutation-based fuzzing studied for decades, while being fully automated, generalizable, and applicable to domains challenging for traditional approaches (such as DL systems). We hope \tech can stimulate more work in this promising direction of LLMs for fuzzing.

\end{abstract}

\maketitle

\section{Introduction}

Deep Learning (DL) is constantly providing revolutionary results and systems in critical fields like autonomous driving~\cite{huval2015autonomousdriving, zhang2018deeproad}, healthcare~\cite{healthcare}, and finance~\cite{finance}. To build these systems, developers use popular DL libraries such as \tf~\cite{Tensorflow} and \pt~\cite{PyTorch} by composing individual library API calls, typically exposed in \python, to build models and perform computations. Due to the significance of detecting and fixing bugs in these DL libraries, researchers have applied various automated bug-finding techniques to test/analyze these libraries~\cite{freefuzz, deeprel, cradle, lemon, audee, eagle, tzer, docter}. One such popular methodology is fuzzing~\cite{SuttonFuzzingBook, Boehme2021fuzzing, zeller2019fuzzing} -- where a large set of inputs are generated and fed to the libraries to find potential bugs. 

Previous work on fuzzing DL libraries mainly falls into two categories: API-level fuzzing~\cite{freefuzz, docter, deeprel} and model-level fuzzing~\cite{cradle, lemon, audee}. API-level fuzzing focuses on testing individual library APIs by generating various different inputs for each target API to discover potential crashes or result inconsistencies. On the other hand, model-level fuzzing techniques aim to generate diverse complete DL models and then compare the model outputs on different backends (e.g., different low-level libraries of Keras~\cite{Keras}) to discover potential bugs. While both model-level and API-level fuzzing techniques have been shown effective in bug finding, they still suffer from the following limitations:

\begin{figure}
    \captionsetup{justification=centering}
    \centering
    \includegraphics[width=\linewidth]{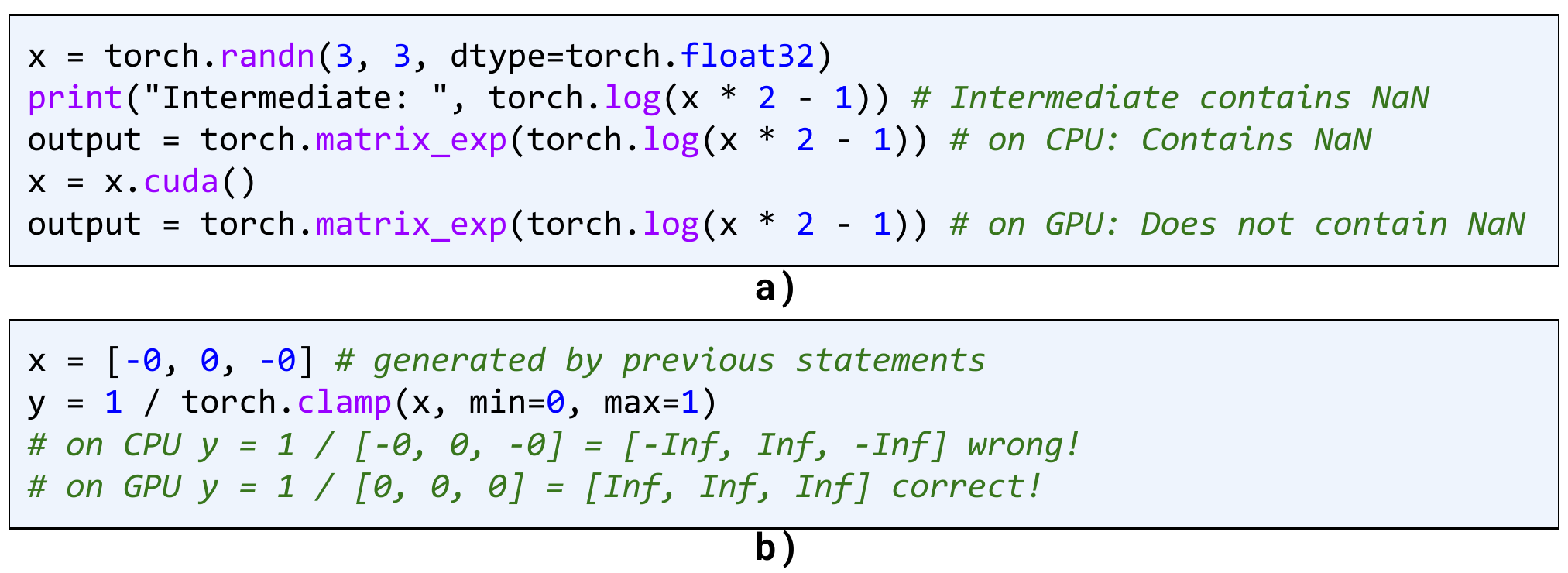}
    \caption{Bugs in DL Libraries}
    \label{fig:intro_example_bugs}
\end{figure}

\textit{1) Lack of diverse API sequences.} Previous API-level fuzzers~\cite{freefuzz, deeprel} only focus on fuzzing each single DL library API in isolation. These techniques attempt to create many different inputs to a particular target API via simple mutation rules. However, these inputs are usually constructed by single code lines (at most one library input creation API, e.g., randomly initializing a tensor with a certain data type and shape) and cannot reveal bugs that are caused by chained API sequences. While model-level fuzzing techniques~\cite{cradle, lemon, audee} can potentially test API sequences, the mutation rules usually have strict constraints, e.g., \lemon's layer addition rule cannot be applied to layers with different input and output shape~\cite{lemon}, while \muffin needs to manually annotate input/output restrictions of considered DL APIs and uses additional reshaping operation to ensure valid connections~\cite{muffin}. As a result, model-level fuzzers can only cover a limited set of APIs with limited patterns~\cite{freefuzz}, missing many diverse and interesting API sequences that can lead to bugs. 

Figure~\ref{fig:intro_example_bugs}a shows an example bug in \pt exposed by an API sequence. A random input is created and the code produces an intermediate variable by invoking the \CodeIn{log} API. The \CodeIn{log} function will produce \CodeIn{NaN} (Not a Number) for negative inputs. In theory, when \CodeIn{matrix\_exp} is applied it should also contain \CodeIn{NaN} values. However, when running this code on GPU, it does not output any \CodeIn{NaN} values. 
More interestingly, this incorrect behavior on GPU cannot be reproduced if we just pass the intermediate tensor (which contains \CodeIn{NaN}) instead of the API call \CodeIn{log} to \CodeIn{matrix\_exp}. %
The bug is only triggered by a synchronization error when we apply the API sequence. Prior API-level fuzzers cannot find this bug; model-level fuzzers can also hardly detect this bug as the specific sequence of \CodeIn{log} followed by \CodeIn{matrix\_exp} is rarely used in building DL models where the focus is on layer APIs such as \CodeIn{Conv2d} or \CodeIn{MaxPool2d}.

\textit{2) Cannot generate arbitrary code.} DL library APIs are exposed to the end user in Python which is not a statically typed language, making it hard to directly obtain the input and output argument types. Also, library APIs usually operate on input tensors where a shape mismatch (e.g., matrix multiplication with incorrect dimensions) can lead to runtime errors. Traditional program synthesis techniques~\cite{manna1971sythesis, solar2008program, odena2020bustle} typically restrict themselves to a small set of functionalities in the language and cannot deal with a large number of library APIs, each with their own specific input and output parameters and types. As such, existing DL library fuzzers use predefined generation grammars that focus on mutating a small part of the program to minimize these errors. %
This limits the variety in both code structure and also the types of inputs that we can use to test library APIs. 
For example, \freefuzz~\cite{freefuzz}, a state-of-the-art API-level fuzzer, will first collect the valid argument space (e.g., type and shape of the input tensor) for a target API by mining open-source code snippets. During the fuzzing loop, \freefuzz will perform small mutations of these valid inputs to generate new inputs, e.g., changing the data type (e.g., \CodeIn{float32} to \CodeIn{float16}). 
As such, \freefuzz is limited by the traced argument space and predefined mutation rules.
Meanwhile, \muffin~\cite{muffin}, a state-of-the-art model-level fuzzer, generates diverse models via using manually annotated specifications for each manipulated API and predefined code structures (e.g. models consisting of sequential layers) in order to preserve model validity.
As such, such prior DL library fuzzers cannot fully explore the huge search space that exists when it comes to using DL library APIs.

Figure~\ref{fig:intro_example_bugs}b shows an example bug in \pt which cannot be detected by previous fuzzing techniques. The bug is caused by the \CodeIn{clamp} function not clamping negative zero to positive zero on CPU. Even though this bug is due to a single API, previous techniques cannot detect this bug as negative zero is \textit{almost} zero. However, the bug is exposed when we apply 1 divided by the clamped list where the correct value should be Positive Inf not Negative Inf (significant value difference). Such \python basic expression is often used by developers in combination with library APIs. However, this bug is missed by prior work due to the restricted generation methods of both existing API-level and model-level techniques.

\mypara{Our Work.} We propose \tech{} -- the first fully automated approach for fuzzing DL libraries via \plmfull{s} (\plm{s})~\cite{brown2020gpt3}. As discussed earlier, DL libraries expose APIs mostly in Python (dynamically typed), making it challenging to directly apply traditional program synthesis to generate syntactically/semantically valid DL programs~\cite{austin2021sythesis}. Moreover, DL APIs may involve complicated input/shape constraints for tensor computations that are extremely hard to satisfy without additional manual efforts. In contrast, modern \plm{s} can serve as a natural solution as they are built using the popular Transformer~\cite{vaswani2017attention} architecture which allows for autoregressive generation (based on left context) or infilling (based on bi-directional context) trained using billions of code tokens to generate ``human-like'' programs. Our key insight is that \emph{modern titanic \plm{s} can include numerous code snippets using various DL libraries in their training corpora (e.g., there are >400,000 TensorFlow/PyTorch projects on GitHub, which is an important training source for modern \plm{s}), allowing them to implicitly learn both Python syntax/semantics and intricate types/constraints of DL APIs to directly generate/mutate valid DL programs for fuzzing DL libraries.}

In \tech{}, we first use a \generative \plm with a step-by-step input prompt~\cite{promptsurvey2022} to produce the initial seed programs for fuzzing. To enrich the pool of test programs, we further adopt an evolutionary strategy to produce new test programs by using \plm{s} to automatically mutate the seed programs. This mutation process is done using multiple mutation operators designed to leverage an \infilling \plm to replace only parts of the seed with new code. In order to generate more complicated and diverse API call relations, we design a fitness function which prioritizes seeds or mutated test programs based on data-dependency depth and number of unique library APIs, allowing us to discover bugs that can only be found when studying complex API relationships. Finally, we execute the generated test programs with differential testing on different backends to detect bugs. In fact, both bugs in Figure~\ref{fig:intro_example_bugs} which cannot be detected by any previous DL library fuzzers are detected by \tech and confirmed by developers as previously unknown bugs. While our approach is general and can be built upon any \plm{s}, we build our technique on \codex~\cite{codex} and \incoder~\cite{incoder} as they have shown state-of-the-art results for generative and infilling tasks, respectively. Also, while we evaluate on two most popular DL libraries: \tf and \pt, our idea of directly using \plm{s} as the generation engine can be applied for fuzzing any DL libraries with little additional effort and can even be extended for fuzzing/testing software systems from other application domains. In summary, this paper makes the following contributions:
{\setlength{\leftmargini}{12pt}
\begin{itemize}
    \item \textbf{Dimension.} This paper opens a new dimension for fuzzing DL libraries (and beyond) by directly using \plm{s} as generation engines. To our knowledge, this is also the first work demonstrating that modern titanic \plm{s} can directly perform both generation-based~\cite{yang2011finding} and mutation-based~\cite{afl} fuzzing studied for decades, while being fully automated, generalizable, and applicable to domains challenging for traditional approaches (such as DL systems). Our approach can be easily extended to test software systems from other application domains (e.g., compilers, interpreters, DB systems, SMT solvers, and other popular libraries). Moreover, this paper demonstrates the promising future of directly leveraging modern \plm{s} for fuzzing and testing in general.  
    \item \textbf{Technique.} We implement \tech, a fully automated fuzzer for DL libraries that first uses a generative \plm (\codex) to synthesize high-quality seed inputs and then combines an infilling \plm (\incoder) with an evolutionary algorithm to guide the generation towards a higher number of unique library API usages and valid/diverse DL programs. 
    \item \textbf{Study.} We perform an extensive evaluation on two of the most popular DL libraries: \pt and \tf. Our result shows that \tech is able to cover \numPtCoverAPI{} / \numTFCoverAPI APIs with \ptCoverage{} / \tfCoverage coverage on \pt and \tf respectively, improving on the state-of-the-art fuzzing tools by \numPtCoverAPIImprove{} / \numTFCoverAPIImprove in API coverage and \ptCoverageImprove{} / \tfCoverageImprove in code coverage. In addition, \tech is able to detect \numTotalBugs bugs, with \numUnknownBugs already confirmed as previously unknown bugs. Furthermore, we perform a broad ablation study to justify the design of components in \tech. %
\end{itemize}}
\section{Background and Related Work}
\subsection{Fuzzing Deep Learning Libraries}

\begin{figure}
    \captionsetup{justification=centering}
    \centering
    \includegraphics[width=0.8\linewidth]{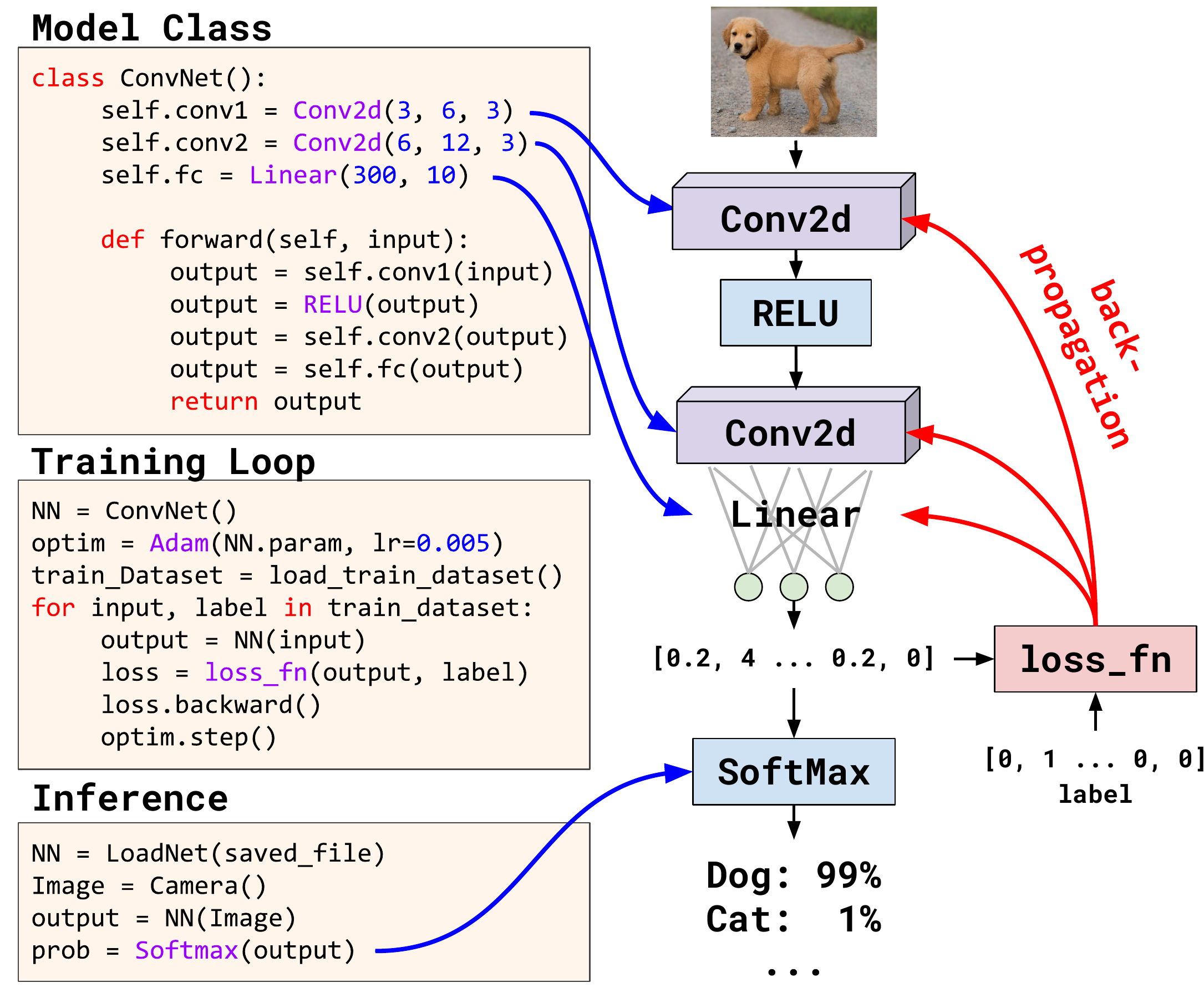}
    \caption{Deep Learning basics}
    \label{fig:dl_background}
\end{figure}

DL libraries (e.g., \tf~\cite{Tensorflow} and \pt~\cite{PyTorch}) serve as the fundamental building block for all DL pipelines by providing thousands of APIs for building, training, and deploying DL models. Figure~\ref{fig:dl_background} shows an example DL model that classifies an input image with its associated training and inference steps. The DL model consists of two convolutional (\CodeIn{Conv2d}) and one fully connected linear (\CodeIn{Linear}) layers. In the forward pass, the first convolutional layer with a non-linear activation function (\CodeIn{RELU}) produces an intermediate output, which is then passed to the second convolutional layer. Next, the fully connected layer is called to produce the final output. In short, using these sets of library APIs, which define the functionality of each layer, the DL libraries essentially create a computational graph, highlighting the flow of data in the model as shown on the right side of the figure. In order to train the model, we first initialize it together with an optimizer that updates the model weights. Next, we load the training data, and for each pair of input and its associated label, we obtain the model output. Finally, we compute the loss together with its gradient to perform back-propagation and update the model weights. To use the model for inference, we first load the trained model and then pass the chosen image to get the model output. Further, we can use the \CodeIn{Softmax} API to obtain the probability representation of the output.

Prior work on fuzzing DL libraries can be mainly classified into two categories, namely model-level and API-level fuzzers. Model-level fuzzers attempt to leverage complete DL models (which cover various sets of DL library APIs) as test inputs. \cradle~\cite{cradle} is one of the first work in the area that detects inconsistencies by running existing models on multiple low-level backends of Keras~\cite{Keras}. To generate more diverse models, \lemon~\cite{lemon} and \audee~\cite{audee} further extend the idea of \cradle to apply predefined mutation rules on seed models/inputs. \muffin~\cite{muffin} further applies a top-down approach to generate DL models for bug detection in both the inference and training phases. Very recently, \nnsmith~\cite{nnsmith} leverages symbolic constraint solving and gradient-based search for high-quality model synthesis. 
While such model-level fuzzers are able to find bugs in DL libraries, 
due to the input/output constraints of DL APIs, model-level mutation/generation rules either are restrictive to certain shape-preserving APIs~\cite{lemon} or require manual annotation of the restrictions of all targeted APIs~\cite{muffin}, leading to a limited number of unique APIs covered. Different from model-level fuzzing, API-level fuzzing focuses on finding bugs within a single API at a time. \freefuzz~\cite{freefuzz} is an API-level fuzzer that first learns the valid inputs for each target API through mining open-source code snippets and then applies simple mutations to generate diverse inputs to test a target API. Similarly, \docter~\cite{docter} mines the input constraints from API documentation by learning the extraction rules with 30\% manually annotated API parameters, and then generates valid and invalid inputs based on the extracted constraints to detect crashes. More recently, \deeprel~\cite{deeprel} and \nablafuzz~\cite{yang2023fuzz} further leverage relational APIs (e.g., APIs that always return the same results/statuses given the same inputs) and automatic differentiation, respectively, as the test oracle for more effective API-level DL library fuzzing. While researchers have demonstrated that API-level fuzzing can cover many more DL library APIs than model-level fuzzing~\cite{freefuzz, deeprel, yang2023fuzz}, API-level fuzzers cannot detect any bug that arises from interactions within a complex API sequence.

\begin{figure}
    \captionsetup{justification=centering}
    \centering
    \includegraphics[width=\linewidth]{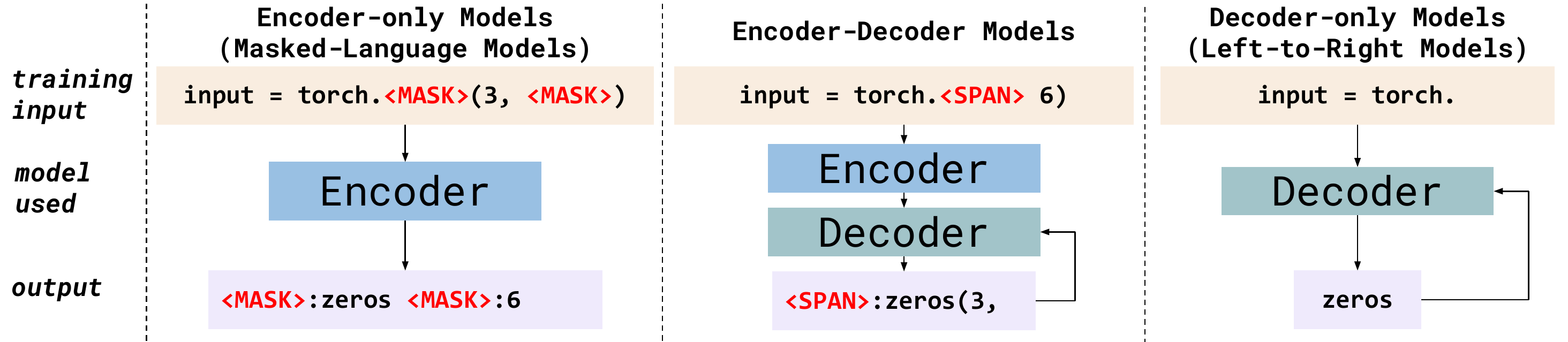}
    \caption{Overview of different \plm architectures}
    \label{fig:plm_background}
\end{figure}

\subsection{\plmfull{s}}

\plmfull{s} (\plm{s}) typically follow the Transformer~\cite{vaswani2017attention} architecture of an \emph{encoder} to produce an encoded representation of the input and a \emph{decoder} to generate output tokens. These \plm{s} are pre-trained on billions of available text on the internet and have been widely used in many different \nlpfull (\nlp) tasks~\cite{brown2020gpt3, liu2019finetune, yang2020xlnet}. Due to the large amounts of available pre-training data, \plm{s} without any fine-tuning on specialized datasets can already be directly used for very specific downstream tasks. This is accomplished using prompt engineering~\cite{promptsurvey2022}, where a natural language description of the task together with a few demonstrations of the task is provided to the \plm first before the actual input. Researchers have demonstrated that this paradigm of directly leveraging \plm{s} through prompts can already achieve state-of-the-art performance on downstream tasks~\cite{brown2020gpt3}. More recently, supported by code naturalness~\cite{hindle2012natural}, researchers have begun to apply \plm{s} for programming languages~\cite{codex, incoder, feng2020codebert, xu2022syseval}. Similar to the impressive performance achieved on tasks in natural languages, \plm{s} can also excel in many code related tasks such as code completion~\cite{codex}, code synthesis~\cite{austin2021sythesis}, and automated program repair~\cite{xia2022alpharepair, xia2022repairstudy}.

Based on the model architectures and pre-training tasks, \plm{s} can be categorized into: \textbf{Decoder-only}, \textbf{Encoder-only} and \textbf{Encoder-decoder}. Figure~\ref{fig:plm_background} shows the three \plm types. Decoder-only models~\cite{brown2020gpt3, codex} use the decoder to predict the probability of the next token based on all previous tokens. These models can be used in an auto-regressive manner to perform auto-completion given all previous (or left) context. Encoder-only models~\cite{feng2020codebert, guo2021graphcodebert} aim to provide a representation of the input through the use of the encoder component. Such models are trained using the Masked Language Model (MLM) objective where a percentage (e.g., 15\%) of the tokens during training are masked out, and the goal is to recover the true values of these masked tokens based on both the context before and after.
Encoder-decoder models~\cite{raffel2020t5, ahmad2021PLBART} use both the encoder and decoder component and are most commonly trained using the Masked Span Prediction (MSP) objective. Instead of replacing each chosen token with a masked token, MSP replaces a sequence of tokens with a single masked span token. The model is then asked to recover the entire sequence during training. During inference, these models can be used to directly \textit{fill in} code in the middle using both the context before and after. However, training both the encoder and decoder components can be time-consuming. As such, recently, researchers have proposed the \incoder~\cite{incoder} model which uses only the decoder component but can fill in text/code in the middle through the Causal Language Model training objective~\cite{incoder}. Instead of using the decoder to autoregressively predict the next token in the original training data, similar to MSP, \incoder also replaces random sequences in the training data with masked span tokens. Using this processed training data, \incoder only autoregressively predicts the original masked sequence given the processed input. With this training strategy, the resulting \incoder model is also able to perform infilling and achieve state-of-the-art results. 

In summary, \plm{s} can perform two main types of code generation tasks: \emph{\generative} and \emph{\infilling}. \Generative tasks involve auto-completing a complete code snippet given the left context only (e.g., some starting code or a natural language description), typically done using decoder-only models. \Infilling tasks aim to insert the most natural code based on bi-directional context (e.g., in the middle of a code snippet), which can be done using encoder-only, encoder-decoder, and also decoder-only models that are trained for infilling, such as \incoder. In this work, we leverage modern \plm{s} to perform both types of generation tasks for fuzzing DL libraries. Besides generative models, inspired by recent work~\cite{xia2022alpharepair, xia2022repairstudy} on \infilling-style program repair (where \plm{s} generate correct patches by directly filling in the correct code given the context), we also leverage \infilling models to perform mutations by replacing a small part of an input program with masked tokens and then filling in generated code to produce even more diverse programs. %

\subsection{Testing using Deep Learning Models}

Due to the recent advances in Deep Learning (DL), researchers have looked into using DL models to facilitate automated test generation or fuzzing of different software systems. 
Traditionally, such techniques rely on training a neural network to produce code snippets automatically.
Seqfuzzer~\cite{seqfuzzer} employs a Recurrent Neural Network (RNN)~\cite{cho-etal-2014-learning, hochreiter1997lstm} and GANFuzz~\cite{ganfuzz} leverages the Generative Adversarial Network (GAN)~\cite{goodfellow2014generative} for protocol fuzzing. Learn{\&}Fuzz~\cite{learnandfuzz}, DeepSmith~\cite{cummins2018compiler} and DeepFuzz~\cite{liu2019deepfuzz} each trains a RNN to generate programs for PDF file parsers, OpenCL and C, respectively. Similarly, Montage~\cite{lee2020montage} targets JavaScript engines by training a RNN to mutate subtrees of a seed input to produce valid JavaScript programs. None of the above work has leveraged \plm{s} for fuzzing.

More recently, COMFORT~\cite{ye2021automated}, has been proposed to fine-tune GPT-2 (with 1.5B parameters)~\cite{radford2019language} on open-source JavaScript programs. COMFORT then uses the fine-tuned GPT-2 model to synthesize JavaScript programs to test specific engines. While COMFORT has demonstrated the potential of \plm{s} for fuzzing, it did not leverage state-of-the-art \plm{s} for code and thus requires an extensive fine-tuning dataset. Moreover, COMFORT cannot perform end-to-end test generation using GPT-2, and has to rely on additional heuristics to generate inputs for the synthesized programs. In contrast, our work demonstrates for the first time that directly leveraging state-of-the-art \plm{s} (e.g., \codex with 12B parameters) can already perform end-to-end input generation for fuzzing real-world systems (without any further fine-tuning). Also, our work shows for the first time that step-by-step prompt engineering~\cite{reynolds2021prompt} can substantially help boost fuzzing. Moreover, to our knowledge, we are the first to apply \infilling models (e.g., \incoder) to directly perform mutation-based fuzzing~\cite{afl, le2014compiler} to generate more diverse input programs in an evolutionary fuzzing loop. 

In addition to using DL techniques for fuzzing, another very recently explored direction involves using \plm{s} for automated unit test generation, e.g., GitHub Copilot has been shown to be promising for such purposes~\cite{copoilottests}. Different from fuzzing which focuses on general approaches for testing complex real-world software systems, unit test generation involves targeting particular modules or functions. As such, unit test generation requires additional knowledge from the program under test such as callable modules (e.g., constructors) and functions. TeCo~\cite{nie2023teco} fine-tunes the CodeT5~\cite{wang2021codet5} model to perform test completion for any targeted method under test and the test signature written by human developers. TestPilot~\cite{schafer2023testpilot} directly uses \codex by prompting with the source code and example usages of the method under test to automatically generate unit tests. Additionally, TestPilot also involves an adaptive component which re-generates failed unit tests by querying \codex given the error message. CodaMosa~\cite{lemieux2023codamosa} combines traditional search-based software testing (SBST)~\cite{fraser2011evosuite, fraser2012whole} with \plm{s} (e.g., \codex) for effective unit test generation. Such existing techniques on \plm-based unit test generation are project/system specific by requiring precise information such as detailed code units under test and/or dynamic test execution traces combined with prompting to elicit generation of unit tests. In contrast, our approach directly leverages the pre-training strategies of \plm{s} to auto-complete/infill code and therefore can easily generalize to arbitrary real-world systems such as compilers/interpreters of different programming languages, DB systems, SMT solvers, and additional popular libraries with sufficient code examples in the massive pre-training corpora of \plm{s}. Furthermore, such unit testing techniques can only assist developers since even current largest \plm{s} (e.g., PaLM~\cite{chowdhery2022palm} and ChatGPT~\cite{chatgpt}) cannot reliably produce oracles for unit tests, while \tech on the other hand is a fully automated approach through the usage of effective fuzzing oracles (such as differential testing) at the system level, and has already detected various bugs for real-world systems. Additionally, \tech is the first work to demonstrate that \plm{s} can perform both generation-based~\cite{yang2011finding, holler2012fuzzing} and mutation-based~\cite{afl, le2014compiler} fuzzing studied for decades~\cite{zeller2019fuzzing}, while being fully automated, generalizable, and applicable to domains challenging for traditional approaches (such as DL systems).

\section{Approach}

\begin{figure*}[t]
    \captionsetup{justification=centering}
    \centering
    \includegraphics[keepaspectratio=true,width=0.85\textwidth]{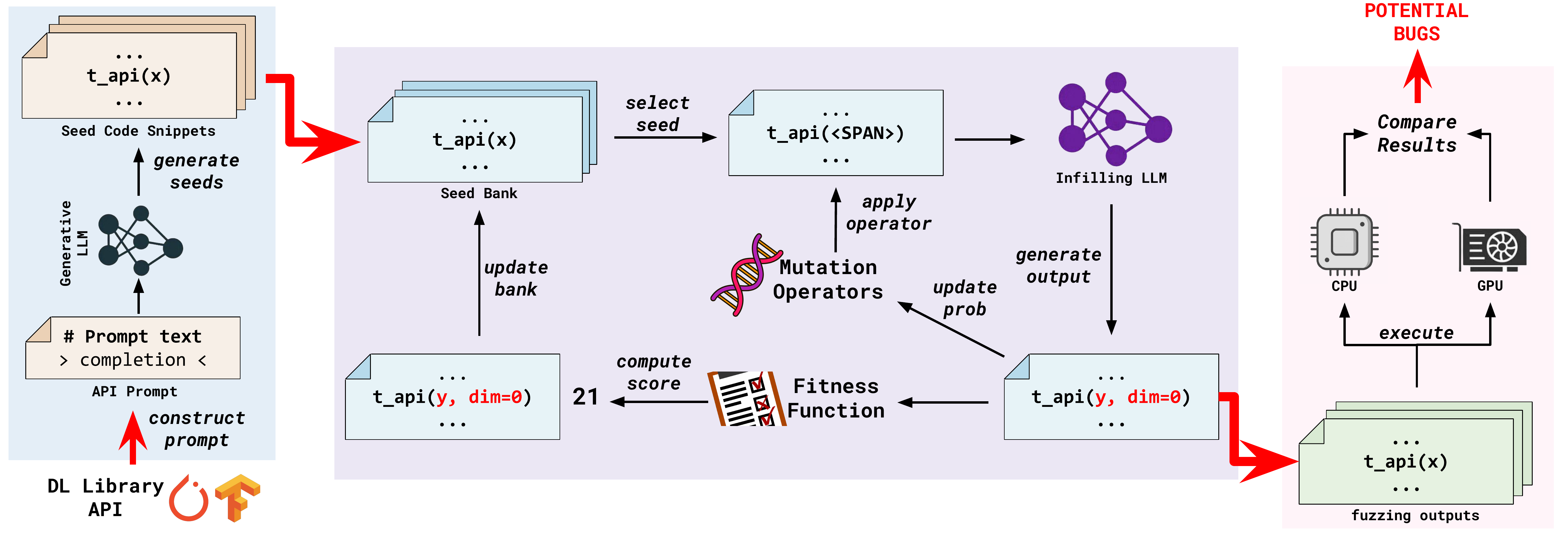}
    \caption{Overview of \tech}
    \label{fig:overview}
\end{figure*}

 Figure~\ref{fig:overview} shows the overview of our \tech approach. Given any target API, \tech first uses a \generative \plm to generate a list of high-quality seed programs for fuzzing (Section~\ref{sec:seed_gen}). This is done by providing the model with a step-by-step prompt~\cite{reynolds2021prompt} to generate code snippets that directly use the target API. For the generated seeds, we further apply an evolutionary fuzzing algorithm to iteratively generate new code snippets (Section~\ref{sec:ev_fuzzing}). In each iteration, we start by selecting a seed program with a high fitness score from the seed bank. We systematically replace parts of the selected seed with masked tokens using different mutation operators (Section~\ref{subsec:mutop}) to produced masked inputs. Mutation operators are selected using a multi-armed bandit algorithm~\cite{thompson1933} (Section~\ref{subsec:mutatorsel}) aiming to maximize the number of valid and unique mutations generated. Using the masked inputs, we leverage the ability of \infilling \plm{s} to perform \textit{code infilling} to generate new code that replaces the masked tokens (Section~\ref{subsec:generation}). For each generated mutant, we first filter out any execution failures and use our fitness function (Section~\ref{subsec:fitfunc}) to score each mutant. We then place all generated mutants into the seed bank, and for future mutation rounds, we prioritize seeds that have a higher score, allowing us to generate a more diverse set of high-quality code snippets for fuzzing.
Finally, we execute all generated programs using differential testing oracle on different backends (CPU and GPU) to identify potential bugs (Section~\ref{sec:oracle}). 

While our approach is general for any pair of \generative and \infilling \plm{s}, in this work, we use \codex~\cite{codex} and \incoder~\cite{incoder} as our \generative and \infilling models, respectively. \codex is a state-of-the-art \generative code model based on the popular GPT-3~\cite{brown2020gpt3} architecture where the model weights are first initialized using GPT-3 weights trained on natural language text and then fine-tuned on a large corpus of open-source code files. \codex can be used to perform auto-completion where the input is simply a description of the task (known as a prompt~\cite{brown2020gpt3,lester-etal-2021-power}). In \tech, we use \codex to automatically create the high-quality seed programs for our evolutionary fuzzing algorithm. 
To obtain mutations from the seed programs, we use the \incoder model to perform \textit{code infilling}. Unlike the \generative models such as \codex 
which only uses the context before, \incoder is able to \textit{fill in} code in the middle by using both the context before and after. In \tech, we combine the power of both types of \plm{s} by first using the \generative model (\codex) to produce high-quality seed programs and then using the \infilling model (\incoder) to generate additional mutated programs. 

\subsection{Initial Seed Generation}
\label{sec:seed_gen}

\begin{figure}
    \captionsetup{justification=centering}
    \centering
    \includegraphics[width=0.9\linewidth]{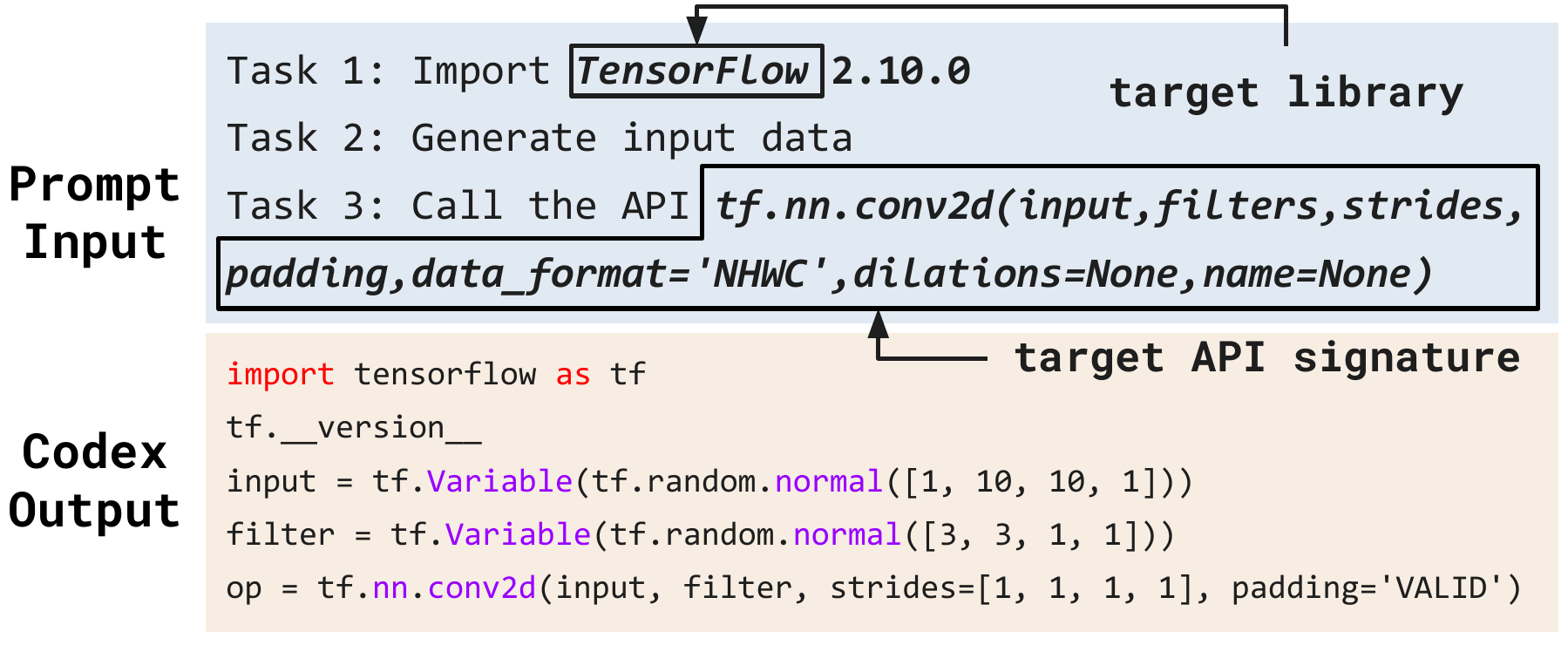}
    \caption{Example generation from the Codex model. }
    \label{fig:example_gen}
\end{figure}

\newcommand{\prefixm}{T_{<m}}
\newcommand{\prefixi}{T_{<i}}
To generate the initial seed programs for a target DL API, we first query the \codex model with a step-by-step prompt and sample multiple completions. \codex is trained using causal language modeling where the model aims to predict the next token using all previous generations. Given a training sequence of tokens $T = \{t_1, t_2, ..., t_n\}$, let $\prefixm = \{t_1, t_2, ..., t_{m-1}\}$ be the token sequence generated so far ($m \leq n$) and $P$ be the \codex model which outputs the probability of generating a token. The \codex loss function is defined as:
\begin{equation}
    \mathcal{L}_{\codex} = -\frac{1}{n}\sum_{i=1}^{n}log\;(P\;(t_i\;|\;\prefixi))
\end{equation}
Figure ~\ref{fig:example_gen} shows an example of the constructed prompt and model output. In the prompt, we wrap our task description in a docstring following ~\cite{codex}. More specifically, we include the target library (e.g., \CodeIn{TensorFlow}) and the target API signature (e.g., \CodeIn{tf.nn.conv2d(input, filters, ...)}) in the prompt. The API signature is automatically extracted from the API documentation with an HTML crawler. We also design a step-by-step instruction (i.e., \CodeIn{Task 1: ... Task 2: ... Task 3: ...} in Figure~\ref{fig:example_gen}) to improve the model's performance following ~\cite{promptsurvey2022,wei2022chain,scratchpad,codexbestpractice}. More precisely, we instruct the model to perform three tasks sequentially: (1) import the target library; (2) generate input data; and (3) invoke the target API. The constructed prompt serves as the initial input for \codex and the raw seed programs are obtained by sampling the autocompletion from \codex.

\subsection{Evolutionary Fuzzing}
\label{sec:ev_fuzzing}
Algorithm~\ref{alg:gafuzz} describes the main evolutionary fuzzing algorithm of \tech. We start by initializing the seed bank with the \codex generated seeds (Line~\ref{lst:line:init}). The seed bank maintains the list of code snippets that have been generated so far. Next, we initialize the prior distribution of each mutation operator, which we will use and update in the main loop for selecting mutation operators (Line~\ref{lst:line:mutprobinit}). We then enter the generation loop by selecting a current seed for mutation according to a fitness score (Line~\ref{lst:line:selectparent}). 
This seed selection process first prioritizes those with higher fitness scores by choosing the top $N$ seeds with the highest fitness score. Out of the top $N$ seeds, we perform a \emph{softmax}~\cite{gibbs1902elementary} operation on their exact fitness scores to determine the probability of picking each seed. %

Besides the seed selection, we also decide which mutation operator will be applied on the seed
(Line~\ref{lst:line:selectmutop}). Since the mutation operators that work well in helping the model to generate valid and unique mutations can be different for different target APIs, we dynamically learn the operator prioritization strategy using a Multi-Armed Bandit (MAB)~\cite{thompson1933} algorithm. 
Each mutation operator will mask out one or multiple segments of the current seed program with a special \CodeIn{<SPAN>} token (Line~\ref{lst:line:genmutinput}). The masked input is then fed into the \incoder model to generate code snippets that infill the masked regions (Line~\ref{lst:line:modelgen}) with sampling. 
For each sample generated, we run and statically analyze the code snippet (Line~\ref{lst:line:eval}). Specifically, we determine the code snippets that can be compiled (\CodeIn{ValidSamples}). We then update the posterior distribution of the mutation operator according to the number of valid and invalid samples it produced (Line~\ref{lst:line:update}). For each valid sample, we compute the fitness score according to our defined fitness function (\CodeIn{FitnessFunction}), which is designed to prioritize a diverse set of seeds that have a high number of unique interactions between different APIs, enabling us to discover more potential bugs. 
Using the fitness score, we add these samples into the seed bank for next iteration of seed selection (Line~\ref{lst:line:addpop}). 
Finally, when the time budget is exhausted, we terminate the generation and return the seed bank, now filled with a number of unique code snippets using the target API. 
Next, we will detail the key components for our algorithm. %

\begin{algorithm}[t!]
\setstretch{0.5}
\caption{Evolutionary fuzzing algorithm}
\label{alg:gafuzz}
\SetKwData{api}{API}
\SetKwData{seeds}{Seeds}
\SetKwData{timeout}{T\_Budget} %
\SetKwData{maxnum}{MaxNum}
\SetKwData{pool}{SeedBank}
\SetKwData{mutprobs}{Mprob}
\SetKwData{curtime}{T\_Elapsed}
\SetKwData{parents}{CurrentSeed} %
\SetKwData{children}{Samples}
\SetKwData{numparent}{P}
\SetKwData{stat}{Stats}
\SetKwData{fitness}{FitnessScore}
\SetKwData{mutators}{MutationOp}
\SetKwData{mutatedin}{MaskedInput}
\SetKwData{validsample}{ValidSamples}
\SetKwData{invalidsample}{InvalidSamples}
\SetKwFunction{eval}{Evaluate}
\SetKwFunction{incodergen}{\incoder}
\SetKwFunction{selectparent}{SelectSeed}
\SetKwFunction{selectmutator}{SelectMutationOp}
\SetKwFunction{selectpool}{SelectSeedBank}
\SetKwFunction{mutate}{Mask}
\SetKwFunction{update}{UpdateFitness}
\SetKwFunction{expand}{ExpandPopulation}
\SetKwFunction{updatemprob}{UpdateMPosterior}
\SetKwFunction{initmprob}{InitializeMPrior}
\SetKwFunction{fitnessfunction}{FitnessFunction}
\SetKwFunction{numsamples}{Count}
\SetKwProg{Fn}{Function}{:}{}
\SetKwFunction{EvoFuzz}{EvoFuzz}
\SetKwInOut{Input}{Input}
\SetKwInOut{Output}{Output}

\Fn{\EvoFuzz{\api, \seeds, \timeout}}{
  \Input{The test target \api, the seed programs \seeds, the time budget \timeout}
  \Output{The generated programs}
  \BlankLine
  \pool $\leftarrow$ \seeds \label{lst:line:init}\\
  \initmprob() \label{lst:line:mutprobinit}\\
  \While{\curtime $\le$ \timeout \label{lst:line:termination}}{
      \parents $\leftarrow$ \selectparent(\pool)\label{lst:line:selectparent}\\
      \mutators $\leftarrow$ \selectmutator() \label{lst:line:selectmutop}\\
      \mutatedin $\leftarrow$ \mutate(\parents, \mutators) \label{lst:line:genmutinput}\\
      \children $\leftarrow$ \incodergen(\mutatedin) \label{lst:line:modelgen}\\
      \validsample, \invalidsample $\leftarrow$ \eval(\children) \label{lst:line:eval}\\
      \updatemprob (\mutators, \numsamples(\validsample), \numsamples(\invalidsample)) \label{lst:line:update}\\
      \fitness $\leftarrow$ \fitnessfunction(\validsample) \\ 
      \pool $\leftarrow$ \pool $\cup$ \validsample \label{lst:line:addpop} 
  }
  
  \algorithmicreturn{ \pool }
}
\end{algorithm}

\subsubsection{Mutation Operators}
\label{subsec:mutop}

\begin{figure}
    \captionsetup{justification=centering}
    \centering
    \includegraphics[width=\linewidth]{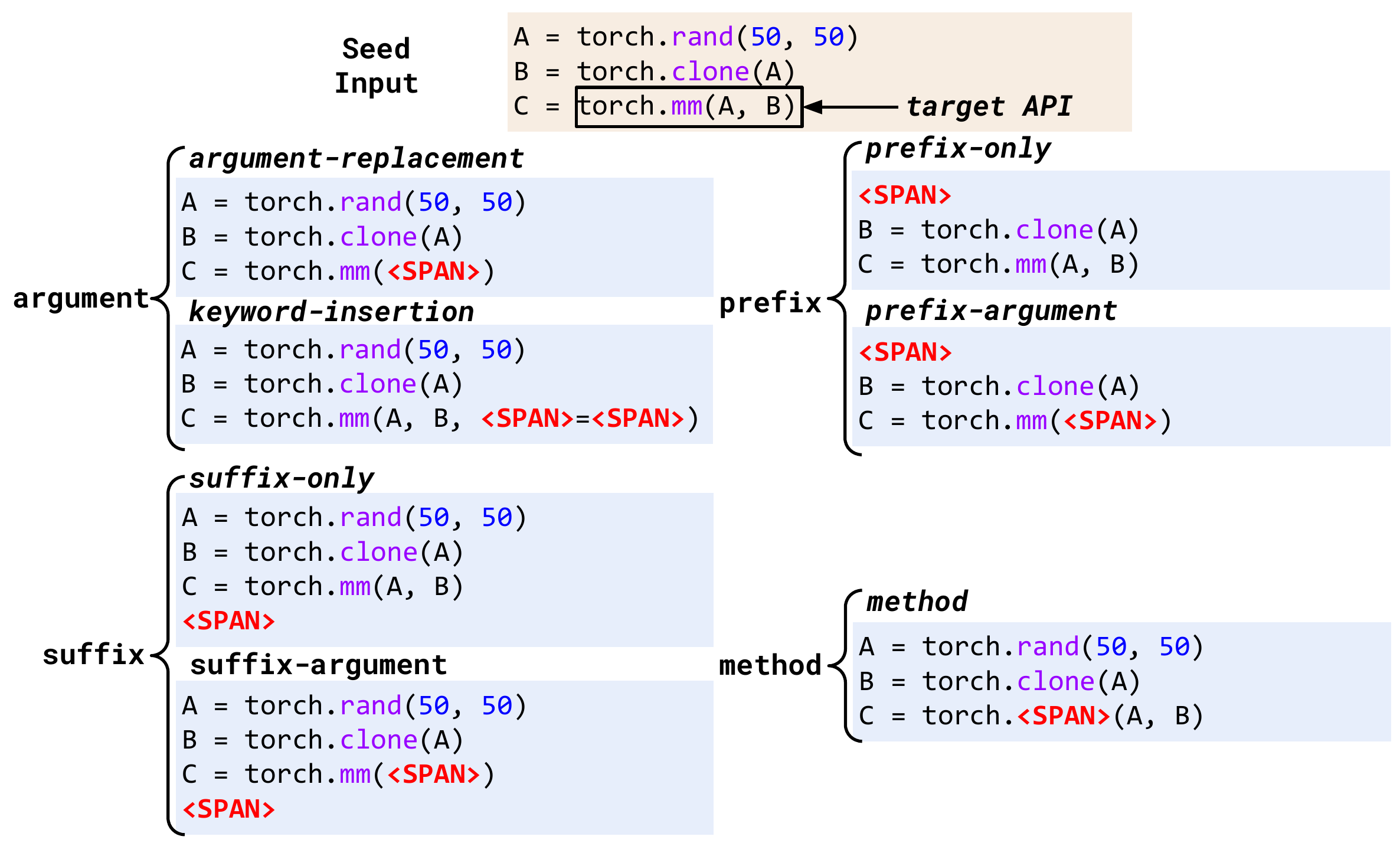}
    \caption{Mutation operators outputs (inputs for the model)}
    \label{fig:example_mut}
\end{figure}

We use four basic types of mutation operators: \textbf{\argument{}, \prefix{}, \suffix{} and \method{}}. Figure~\ref{fig:example_mut} shows the example masked inputs generated using each of our mutation operators. We start by identifying the target API (e.g., \CodeIn{torch.mm}) in the seed code snippet. Each mutation operator replaces a particular location of the chosen seed code with a masked span token (\CodeIn{<SPAN>}). For example, the \argumentmut mutation operator will replace the argument of the target API call with the span token. The key idea is to create inputs which leverage the ability of \plm{s} to generate code snippets which replace each span token at the desired location, i.e., replacing the \CodeIn{<SPAN>} token with the model generation. We now define each of our mutation operator types:

\mypara{Argument}. The first is the \argumentmut mutation operator which replaces the API call arguments with the span token. Using this masked input, the model can fill-in different arguments to the API, generating unique and different program behaviors. Note, the \argumentmut mutation operator is not limited to just the target API and can be applied on any arbitrary library API in the code snippet. Furthermore, we use the \keyword mutation operator which attempts to allow the model to generate additional keywords for a particular library API. Different from the \argumentmut operator which replaces the entire argument list (which can include more than one argument), the \keyword{} operator appends two span tokens at the end of the argument list. The two span tokens sandwich an equal sign, which signifies to the model that the two spans together should form a keyword. The equal sign is important, as without it, the model may think that an additional argument should be added here instead of a keyword. Keywords can unveil many interesting bugs because typically only the base case keyword values are tested; this is especially true for API sequences, as the combination of keywords in multiple API calls can lead to previously undiscovered bugs. 

\mypara{Prefix/Suffix}. Just modifying the arguments of a specific API can only produce limited code mutants, since there are only a limited number of variables/literals available in the current scope for each argument in the API call. As such, to further augment the seed input, we apply \prefix and \suffix mutation operators, which choose a code segment (spanning one or multiple lines) before or after the target API invocation to insert the span token. In Figure~\ref{fig:example_mut}, the \prefixonlymut operator replaces the first line in the seed input with the span token. The model may then fill in this span token with different input generation methods instead of \CodeIn{torch.rand}. On the other hand, the \suffixonlymut operator adds a span token after the target API. This allows the model to generate more code, potentially applying other DL APIs on the output of the prior code, covering additional interesting program behaviors/bugs. Since both the prefix and suffix can affect arguments of the API, we further combine the prefix and suffix together with the argument (i.e., the \prefixargmut and \suffixargmut operators) to allow more freedom for generating code before/after at the same time as argument generation. 

\mypara{Method}. We include the \method operator which replaces a randomly chosen library API method name with a span token. The idea is to generate a different library API invocation while using the existing arguments. 
The inspiration comes from prior work~\cite{deeprel}, which shows that it is common in DL libraries for related APIs to share the same input, and \textit{borrowing} inputs from one API can help trigger bugs in its relational APIs. We can also leverage the \method operator to generate more unique code snippets and test more APIs.

\subsubsection{Mutation Operator Selection}
\label{subsec:mutatorsel}

\begin{algorithm}[t!]
\setstretch{0.5}
\caption{Mutation operator selection algorithm}
\label{alg:mutation_selection}
\SetKwData{mutprobs}{Mprob}
\SetKwData{mutators}{MutationOps}
\SetKwData{mutator}{m}
\SetKwInOut{Input}{Input}
\SetKwInOut{Output}{Output}
\SetKwProg{Fn}{Function}{:}{}
\SetKwFunction{random}{Random}
\SetKwFunction{randomselec}{RandomSelect}
\SetKwFunction{argmax}{argmax}
\SetKwFunction{init}{InitializeMPrior}
\SetKwFunction{selectmutator}{SelectMutationOp}
\SetKwFunction{updatemprob}{UpdateMPosterior}
\SetKwData{stats}{Stats}
\SetKwData{stat}{Stat}
\SetKwData{success}{S}
\SetKwData{failure}{F}
\SetKwData{numvalidsample}{NumValid}
\SetKwData{numinvalidsample}{NumInvalid}
\Fn{\init{}}{
    \For{\mutator $\in$ \mutators \label{ts:line:init} } {
        \mutator.\success, \mutator.\failure $\leftarrow$ 1, 1 \label{ts:line:initendline}
    }
}
\Fn{\selectmutator{}}{
    \Output{The chosen mutation operator \mutator}
    \BlankLine
    \For{\mutator $\in$ \mutators} {
        Sample $\theta_{\mutator}$ $\sim$ Beta(\mutator.\success, \mutator.\failure) \label{ts:line:sample}
    }
    $\mutator^{\star}$ = $arg max_{\mutator} \theta_{\mutator}$ \\
    \algorithmicreturn{ $\mutator^{\star}$}
}

\Fn{\updatemprob{\mutator, \numvalidsample, \numinvalidsample}}{
    \BlankLine
    \mutator.\success $\leftarrow$ \mutator.\success + \numvalidsample \label{ts:upd} \\
    \mutator.\failure $\leftarrow$ \mutator.\failure + \numinvalidsample \label{ts:upd:lastline}
}
\end{algorithm}

\newcommand{\bestarm}{$i^{\star}$}
\newcommand{\predarm}{$\hat{i}$}
\newcommand{\budget}{$n$}
\newcommand{\mab}{MAB\xspace}
\newcommand{\mablong}{Multi-Armed Bandits\xspace}
\newcommand{\armNum}{$K$}
 We first formulate our mutation operator selection problem as a multi-arm bandit (MAB) problem~\cite{thompson1933}, and then detail our algorithm.
 Our assumption is that the mutation operator that works well (e.g., generating more bug-triggering mutants) can be different for different DL APIs. Thus, we would like to adaptively learn the effectiveness of each operator in the generation loop. Intuitively, the validity of generated programs from a mutation operator can be a strong hint for prioritizing/de-prioritizing the operator. Thus, we model the mutation operator selection problem as a Bernoulli bandit problem~\cite{russo2018tutorial} as follows:

\begin{definition} 
\parabf{Bernoulli Bandit.} Suppose there are \armNum{} arms, and when played, each arm yields either a success or a failure. Each arm $i \in \{1,\ldots,\armNum{}\}$ is associated with a success probability $\mu_i \in [0,1]$, which is unknown to the agent.
At each time step $t$, the agent will pull an arm $i_t$ and observe a success/failure output drawn from the Bernoulli distribution $Ber(\mu_{i_t})$. The objective is to maximize the accumulated number of successes over $T$ rounds of experimentation.
\end{definition}

\begin{definition}
    \parabf{Beta-Bernoulli Bandit.}
    For a Bernoulli bandit problem, let the agent adopt a Bayesian framework and choose the standard beta distribution~\cite{betadistribution} as the independent prior belief over each arm $m$. The probability density function of the beta distribution, for $0 \leq x \leq 1$, and parameters $\alpha > 0, \beta > 0$ is given by $$f(x;\alpha,\beta)=\frac{\Gamma(\alpha+\beta)}{\Gamma(\alpha)\Gamma(\beta)}x^{\alpha-1}(1-x)^{\beta-1}$$
    , where $\Gamma$ denotes the gamma function~\cite{gammafunction}. If the prior is a $Beta(\alpha, \beta)$ distribution, the posterior will also be a beta distribution, with $\alpha$ or $\beta$ increases by one with each observed success or failure, respectively.
\end{definition}

For \tech, each mutation operator $m$ can be seen as an arm associated with an unknown expected success probability, defined as the unique pass rate (percentage of generated code snippets when using a mutation operator that are valid and different from historical generations). 
 When we``play'' an arm at time $t$, we apply the mutation operator to generate programs, validate them, and interpret each program execution status as a success or failure.

To balance the exploitation and exploration trade-off in this beta-Bernoulli bandit problem, we leverage the classic Thompson Sampling (TS) algorithm~\cite{thompson1933,tsNIPS2011}.
Algorithm~\ref{alg:mutation_selection} shows how TS, specialized for our mutation operator selection, proceeds. 
By initializing $m.S$ and $m.F$ to 1 (Lines~\ref{ts:line:init}-\ref{ts:line:initendline}), the algorithm assumes each arm $m$ has prior $Beta(1,1)$ (i.e., uniform distribution). After observing $m.S-1$ successes and $m.F-1$ failures of arm $m$, the posterior distribution of $\mu_{m}$ is updated as $Beta(m.S, m.F)$. To select an arm, we draw a sample $\theta_m$ from each of those posterior distributions (Line ~\ref{ts:line:sample}) and play the arm with the largest sampled value (which indicates that it has the highest probability of having the highest success rate). After generation using \plm{s}, we then update the posterior of the chosen mutation operator based on the execution statuses of generated programs (Lines~\ref{ts:upd}-\ref{ts:upd:lastline}). Compared to randomly picking mutation operators to use, this approach allows us to identify mutation operators that help generate more valid and unique code snippets. 
Note the \textit{best} mutation operators can be different for different target APIs, therefore we start a separate \mab game and re-initialize the operator prior distribution for each end-to-end run of the evolutionary fuzzing targeting one API.

\subsubsection{Code Generation}
\label{subsec:generation}

After using the selected mutation operator to produce the masked input for code generation, we use \incoder to generate new code to fill-in the masked-out location. \incoder is trained using causal masking objective~\cite{incoder} {to} perform \textit{code infilling} by using bi-directional context to determine reasonable code snippets to place in the middle. Let $T_{masked} = \{T_1, T_2, ... \CodeIn{<SPAN>}, T_n\}$ be training code tokens where span mask tokens are inserted, $M = \{m_1, m_2, ..., m_k\}$ be the tokens masked out, $M_{<g} = \{m_{1}, m_{2}, ..., m_{g-1}\}$ be the list of tokens generated so far ($g \leq k$), $P$ be the \incoder model which outputs the probabilities of generating a token. The loss function of \incoder can described as:
\begin{equation}
    \mathcal{L}_{\incoder} = -\frac{1}{k}\sum_{i=1}^{k}log\;(P\;(m_i\;|\;T_{masked},\;M_{<i}))
\end{equation}
We leverage the ability of \incoder to generate arbitrary but related code snippets to target intricate library API relationships as the model can learn from the context (surrounding code which already focuses on library APIs) to generate additional APIs/code. As DL library APIs operate on Tensors, a shape mismatch (e.g., vector addition with incorrect dimensions) can lead to runtime errors. Traditional mutations (e.g., changing random code elements) do not work in generating valid DL programs since they can easily cause runtime errors by incorrectly mutating the input and argument space, and also cannot easily ensure the semantic validity~\cite{park2021generative, lampropoulos2019coverage} of the generated programs due to the dynamically typed language. In contrast, \incoder is trained on millions of code snippets, many of which contain usages of these library APIs~\cite{hindle2012natural}. This allows \incoder to directly provide interesting/correct code based on the bi-directional context and generate potentially valid DL programs. Using the code generated by \incoder, we place them directly into the \CodeIn{<SPAN>} of the mutated input to produce new code snippets.

\subsubsection{Fitness Function}
\label{subsec:fitfunc} 
Similar to the mutation operator, the seed programs we choose to mutate over are also important to generate unique and interesting code snippets for fuzzing. As such, we design a fitness function and score to rank each generated program. We apply static analysis to calculate the fitness scores of each test program. The intuition behind the fitness calculation is to give higher scores to the generated mutation programs with deeper execution path, more diverse computation graph, and more complicated API invocations. Specifically, we consider the following features:
{\setlength{\leftmargini}{12pt}
\begin{itemize}
    \item \textbf{Depth of dataflow graph.} We statically analyse the dataflow of variables within the generate code snippets to build a dataflow graph with each edge representing data dependencies between two operations. The depth of the dataflow graph (\textbf{\dagdepth}) is defined as the maximum number of edges in any path of the graph.  
    \item \textbf{Number of API calls.} We count the number of unique library API calls (\textbf{\apicall}) that exist within each code snippets. Since \plm tends to generate many code snippets where code line(s) are repeated, we also count and penalize the number of library APIs that are repeatedly called with the same inputs (\textbf{\repeatedapicall}).
\end{itemize}}

Combining all these factors, given generated code snippet $C$, we define our fitness function to be:
\begin{equation}
    \label{eq:fitnessfunction}
    \textbf{FitnessFunction($C$)} = \textbf{\dagdepth} + \textbf{\apicall} - \textbf{\repeatedapicall}
\end{equation}
According to the formula, \tech favors input programs involving long-chained API sequences and more unique APIs. In this way, it allows us to cover more interactions between different APIs, potentially triggering more interesting program behaviors/bugs. Meanwhile, using only the first two sub-terms would cause \tech to cover longer and longer API sequences with repeated API calls, making the fuzzing process less efficient. Therefore, \tech further penalizes the sequences with repeated API calls.

\subsection{Oracle}
\label{sec:oracle}

After the generation loop, we leverage differential testing oracle to detect bugs by running the generated code snippets on two separate backends. In short, we execute the generated code snippets on CPU and GPU, record all the variables including the intermediate ones, and detect potential bugs. We focus on the following bug types:

\mypara{Wrong-Computation.} We compare the values of all intermediate variables across the two execution backends and find wrong-computation when values are significantly different. Due to the non-deterministic nature of certain computations leading to slightly different results on CPU or GPU, we follow previous work~\cite{freefuzz} and use a tolerance threshold to check if values are significantly different. Difference in computed values can indicate a potential semantic bug in different backend implementations of a library API or interactions between different APIs. 

\mypara{Crash.} During program execution, we also detect unexpected crashes, e.g. segmentation faults, aborts, \CodeIn{INTERNAL\_ASSERT\_FAILED} errors. Such crashes indicate failures to check or handle invalid inputs or corner cases, and can lead to security risks.

\section{Evaluation}

We aim to investigate the following research questions:
\begin{itemize}
    \item \textbf{RQ1:} How does \tech{} compare against existing DL library fuzzers?
    \item \textbf{RQ2:} How do the key components of \tech{} contribute to its effectiveness?
    \item \textbf{RQ3:} Is \tech{} able to detect real-world bugs?
\end{itemize}

\subsection{Implementation}
\label{sec:implementation}

For seed generation, we use the \codex Completion model with \CodeIn{code-davinci-002} engine to sample $25$ programs for each API. Since the \codex model is not open-sourced, we access it by interacting with the \codex API through HTTP requests from Python. Our default setting for code completion for \codex uses top-$p$ (nucleus) sampling~\cite{holtzman2019nucleus} with $p$ = 0.95 following previous studies~\cite{incoder,codex}, and $max\_tokens$=256, $temperature$=0.4 tuned for our task. Since we pose a maximum token limit to \codex model for code completion, the generated program can end with an incomplete line. Thus, for each \codex-produced program, we iteratively remove the last line of the program until the syntax parsing succeeds. For fuzzing, we choose $N$=10 for seed selection and use the \pt implementation of the \incoder 1.3B model on Hugging Face~\cite{HuggingFaceWebPage}. Our default setting when using \incoder uses temperature = 1 with default settings of top $p$ = 0.95 from previous studies~\cite{incoder}. We apply code filtering to remove unnecessary code generated by the model such as print statements. 
Furthermore, we apply dataflow analysis to perform dead code elimination.

\subsection{Experimental Setup}
\parabf{Targeted DL libraries. } We include both \pt (v1.12)~\cite{PyTorch} and \tf (v2.10)~\cite{Tensorflow}, since they are two of the most popular DL libraries and are widely studied in prior DL library testing work~\cite{freefuzz, docter, eagle}.  

\parabf{Fuzzing budget. } By default, we use a one-minute fuzzing budget per API for all possible APIs of both studied libraries. Meanwhile, for RQ2, we randomly sample 100 public APIs in each library and conduct the ablation study experiments for five times and report the average following prior work~\cite{evalfuzz}. Also, for RQ3, we extend the fuzz budget to four-minute per API for maximal bug finding.

\parabf{Environment. } We{ use} a 64-core workstation with 256 GB RAM and running Ubuntu 20.04.5 LTS with 4 NVIDIA RTX A6000 GPUs. We use the coverage.py~\cite{coverage-py} tool to measure Python code coverage.

\subsection{Metrics}

\parabf{Number of detected bugs.} Following prior work on fuzzing DL libraries ~\cite{cradle,lemon,freefuzz,docter,eagle}, we report the number of detected bugs.

\parabf{Code coverage.} Code coverage has been widely adopted in software testing and recently DL library/compiler testing~\cite{freefuzz, muffin, tzer}. We follow recent DL library fuzzing work (Muffin~\cite{muffin} and FreeFuzz~\cite{freefuzz}) and use line coverage.

\parabf{Number of covered APIs.} Following prior work ~\cite{freefuzz,deeprel}, we report the number of covered APIs as another important metric of test adequacy in DL libraries which typically have thousands of APIs.

\parabf{Number of unique valid programs generated.}
A generated program is considered valid if the program executes successfully without exceptions and actually invokes the target API at least once. We also remove the code snippets that have already been generated and only consider unique programs. %

\parabf{Execution time.} Since \tech uses \plm{s} as the generation engines, it may take more time than existing fuzzers. As such, we also record the execution time following prior work~\cite{lemon, freefuzz, tzer}.

\section{Result Analysis}

\subsection{RQ1: Comparison with Prior Work}

We compare \tech against both state-of-the-art API-level (\freefuzz~\cite{freefuzz}, \deeprel~\cite{deeprel}) and model-level (\lemon~\cite{lemon}, \muffin~\cite{muffin}) fuzzers for testing DL libraries. Table~\ref{tab:apicov} presents the number of library APIs covered by all studied techniques on \tf{} and \pt. We run each tool with its default setting, and since \lemon and \muffin do not support \pt models, we only report their results on \tf. Column \textbf{Total} presents the total number of APIs in each DL library. Note that we excluded the deprecated APIs and compatibility APIs in \tf as they are no longer actively maintained by developers.

We observe that \tech is able to cover \numTFCoverAPI and \numPtCoverAPI APIs in \tf and \pt, \textit{achieving the highest number of APIs covered compared to state-of-the-art techniques}. \tech increases the number of APIs covered by \numTFCoverAPIImprove and \numPtCoverAPIImprove compared to the best-performing baseline \deeprel. Compared with model-level fuzzing techniques (\lemon and \muffin), \plm can greatly outperform them in terms of the number of covered APIs. This is due to the fact that model-level fuzzers use complete DL models that are implemented using a small set of layer-wise APIs such as \CodeIn{Conv2d}. On the other hand, \tech is able to generate arbitrary code through the use of both \generative (\codex) and \infilling (\incoder) \plm{s} to achieve state-of-the-art results in terms of API coverage.

\begin{table}[!htp]\centering
\caption{Comparison on API coverage}\label{tab:apicov}
\scriptsize

\begin{tabular}{l|r|rr|rr|rr}\toprule
&\tech{} &\deeprel &\freefuzz &\muffin &\lemon &Total \\\midrule
\tf &\numTFCoverAPI &1159 &581 &79 &35 &\numTFTotalAPI \\
\pt &\numPtCoverAPI &1071 &468 &- &- &\numPtTotalAPI \\
\bottomrule
\end{tabular}

\end{table}

\begin{figure}
    \captionsetup{justification=centering}
    \centering
    \includegraphics[width=\linewidth]{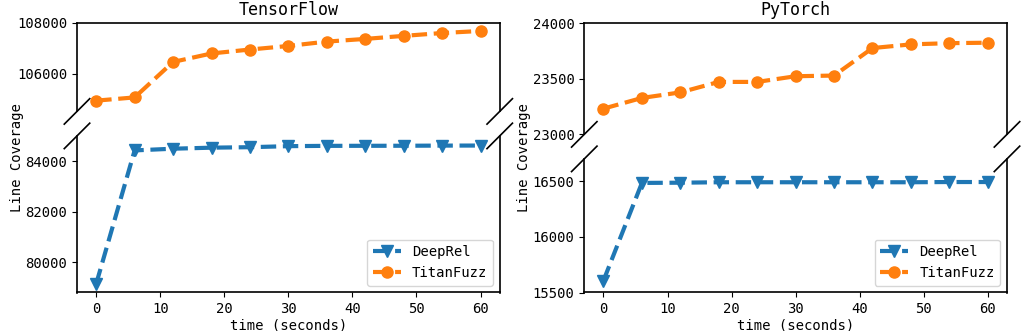}
    \caption{Coverage trend against \deeprel}
    \label{fig:coverage-trend}
\end{figure}

Table~\ref{tab:comp-freefuzz} presents the overall code coverage rate. We choose the best-performing API-level and model-level baselines \deeprel and \muffin according to API coverage and run with their default settings. 
We observe that \tech significantly outperforms both \deeprel and \muffin, \textit{achieving the state-of-the-art result of \ptCoverage and \tfCoverage line coverage on \pt and \tf}. Compared with \deeprel, we increase the coverage result by \ptCoverageImprove and \tfCoverageImprove on \pt and \tf. The time cost of \tech is higher due to the larger number of tested APIs and the use of \plm{s}. However, we observe that simply running \tech with only seed generation and targeting only the APIs that are covered by \deeprel (Row \textbf{\tech{}-seed-only (w/ \deeprel APIs)}) can already greatly outperform \deeprel with even less time, showing the power of directly using \plm{s} to produce high-quality seeds.  

Figure~\ref{fig:coverage-trend} further shows the coverage trend of \tech against the best baseline \deeprel as we increase the time spent on fuzzing each target API. In this experiment, we run both techniques with a one-minute time budget for each API. We note that the \deeprel coverage barely increases after around 10 to 20 seconds of fuzzing. On the other hand, \tech does not suffer from the same coverage saturation. We observe \tech can still generate new programs that improve coverage even after 50 seconds of fuzzing. We attribute this to both the usage of \plm to perform infilling and our \textit{guided} seed and mutation operator selection in the generation process.

\begin{table}[!htp]\centering
\caption{Comparison with the best existing techniques}\label{tab:comp-freefuzz}
\scriptsize

\begin{tabular}{l|rr|rrr}\toprule
&\multicolumn{2}{c|}{\pt} &\multicolumn{2}{c}{\tf} \\\cmidrule{2-5}
&Coverage &Time &Coverage &Time \\\midrule
\deeprel &{15794 (13.91\%)} &5.1h &{82592 (30.65\%)} &9.3h \\ \midrule
\muffin & - & - & {79283 (29.42\%)} & 6.8h \\ \midrule
\makecell[l]{\tech{}-seed-only \\(w/ \deeprel APIs)}&{18447 (16.25\%)} &3.4h &{89048 (33.05\%)} &4.9h \\
\makecell[l]{\tech-seed-only \\(w/ all APIs)}  & {22584 (19.89\%)} &5.1h &{103054 (38.35\%)} &11.9h \\
\tech{} &23823 (20.98\%) &9.9h &107685 (39.97\%) &21.1h \\
\bottomrule
\end{tabular}
\end{table}

\subsection{RQ2: Evaluation of Key Components}

\subsubsection{Seed Generation}

We first study the various design choices for our seed generation which uses the \codex model with a carefully designed input prompt. The goal of the seed generation is to provide high-quality programs for as many APIs as possible. Therefore, we compare several variants of the input prompt and different \codex model hyperparameter values.
 
\begin{figure}
    \captionsetup{justification=centering}
    \centering
    \includegraphics[width=\linewidth]{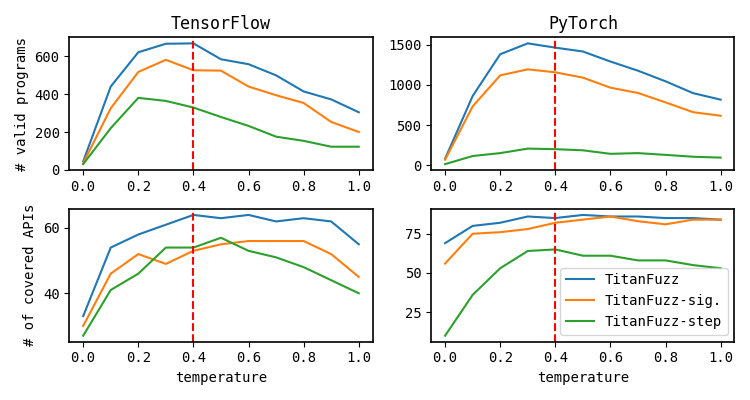}
    \caption{\codex seed generation trend}
    \label{fig:codexseedgen-trend}
\end{figure}

Figure~\ref{fig:codexseedgen-trend} shows the API coverage and number of unique valid programs with different temperatures and prompts. \textbf{\tech} represents the default strategy presented in Section ~\ref{sec:seed_gen}, where the prompt includes three steps to first import the DL library, generate input, and then call the target API. We also include the full API signature in the prompt to provide syntax guidance. \textbf{\tech-sig.} only provides the API name instead of API signature in the prompt (e.g. replacing the entire target API signature with \CodeIn{tf.nn.conv2d} in Figure~\ref{fig:example_gen}). \textbf{\tech-step} removes the first two steps (import library and input generation) and only asks \codex to call the target API from the specified library version. First, we find that our default sampling temperature value of 0.4 (red dotted line) provides a good balance in terms of generating more valid programs and also covering more unique APIs on both \pt and \tf. Second, we observe that by adding step-by-step instructions (first import the library and generate input data) to the prompt, we can substantially improve \codex's performance in both the number of unique valid programs generated and API coverage, demonstrating the power of prompt engineering for fuzzing for the first time. Furthermore, by adding the API signature to the prompt, we provide \codex with valuable information regarding the input parameter space to help \codex generate more valid programs. %

\subsubsection{Evolutionary Generation} 
Next we look at the different factors and choices in the evolutionary fuzzing algorithm.

\parabf{Mutation Operators.} We examine the effectiveness of each of our additional mutation operator types (except the default argument operators). Table~\ref{tab:study-opset} shows results when we remove each type from \tech. Column \textbf{Valid} considers only unique valid programs and Column \textbf{All} also includes programs with runtime errors. We observe that the highest number of unique programs and coverage is obtained when we use full set of mutation operator. This shows that each of the mutation operators can help in producing more unique programs and covering additional lines. 

\begin{table}[!htp]\centering
\caption{Ablation study of operators}\label{tab:study-opset}
\scriptsize
\begin{tabular}{l|rrrr|rrrrr}\toprule
\multirow{3}{*}{Variants} &\multicolumn{4}{c|}{\pt} &\multicolumn{4}{c}{\tf} \\\cmidrule{2-9}
&\multicolumn{2}{c}{{\# Unique Prog.}} &\multicolumn{2}{c|}{{Coverage}} &\multicolumn{2}{c}{{\# Unique Prog.}} &\multicolumn{2}{c}{{Coverage}} \\\cmidrule{2-9}
&{Valid} &{All} &{Valid} &{All} &{Valid} &{All} &{Valid} &{All} \\\midrule
\textbf{\tech} &\textbf{6969} &\textbf{18245} &\textbf{17411} &\textbf{17957} &\textbf{5173} &\textbf{16865} &\textbf{84447} &\textbf{86536} \\
-Suffix &5770 &15813 &16709 &17691 &4642 &14501 &81145 &85294 \\
-Method &6239 &16943 &16886 &17615 &3492 &12519 &83405 &85454 \\
-Prefix &6211 &17082 &17075 &17797 &3359 &12345 &83435 &85645 \\
\bottomrule
\end{tabular}
\end{table}

\parabf{Fitness Function. } We compare our default fitness function against its variants, as well as random selection (\textbf{\randombaseline}) and a simplistic coverage guided~\cite{afl, aflplusplus, libfuzzer, lemieux2018fairfuzz} (\textbf{\coveragebaseline}) baseline in Table~\ref{tab:study-fitness}. The fitness function variants are constructed by removing each sub-term from the original fitness function (Equation~\ref{eq:fitnessfunction}). 
We observe that our chosen fitness function (\textbf{D+U-R}) is able to achieve close to the highest coverage and number of unique programs generated for both \tf and \pt. Compared to random selection, our chosen fitness function approach is able to obtain higher coverage. This is due to the fitness function's ability to \textit{guide} the fuzzing process towards using seeds with more unique APIs and longer chained API sequences, leading to covering more lines of code. Compared to our coverage-guided baseline which only adds programs with new coverage to the seed bank for later mutation, our fitness function has minimal additional overhead. This allows \tech to spend more time on the generation, leading to not only higher coverage but also more unique code snippets for testing.

\begin{table}[!htp]\centering
\caption{Ablation study of fitness function}\label{tab:study-fitness}
\scriptsize

\begin{tabular}{l|rrrr|rrrrr}\toprule
\multirow{3}{*}{{Variants}} &\multicolumn{4}{c|}{{PyTorch}} &\multicolumn{4}{c}{{Tensorflow}} \\\cmidrule{2-9}
&\multicolumn{2}{c}{{\# Unique Prog.}} &\multicolumn{2}{c|}{{Coverage}} &\multicolumn{2}{c}{{\# Unique Prog.}} &\multicolumn{2}{c}{{Coverage}} \\\cmidrule{2-9}
&{Valid} &{All} &{Valid} &{All} &{Valid} &{All} &{Valid} &{All} \\\midrule
D+U-R &\textbf{6960} &\textbf{18245} &17411 &17957 &\textbf{5173} &\textbf{16865} &\textbf{84447} &\textbf{86536} \\
D+U &5817 &15609 &\textbf{17725} &\textbf{18415} &2993 &11253 &82963 &85455 \\
D-R &5872 &16916 &17229 &18046 &2876 &11861 &83563 &85599 \\
U-R &6234 &17321 &16894 &17820 &4315 &15495 &84057 &86286 \\
\hline
\randombaseline &7288 &20720 &16674 &17586 &3274 &13237 &83440 &85045 \\
\coveragebaseline &5098 &15300 &16715 &17617 &3210 &12880 &83030 &84194 \\
\bottomrule
\end{tabular}
\end{table}

\parabf{Operator Selection Algorithm.} We compare our default Thompson Sampling operator selection algorithm (\textbf{TS}) with a uniformly random selection baseline (\textbf{\randombaseline}). Table ~\ref{tab:study-mab } summarizes the results. The TS bandit algorithm helps to generate more unique valid programs and achieve higher code coverage in both libraries compared to the random strategy. Specifically, the TS strategy can generate around 2X more valid unique programs for \tf; in \pt, although TS can generate fewer unique programs in total, it can still produce 12.5\% more valid ones, demonstrating the effectiveness of our \mab-based operator prioritization.
\begin{table}[!htp]\centering
\caption{Evaluation of operator selection algorithms}\label{tab:study-mab }
\scriptsize
\begin{tabular}{l|l|rrrrr}\toprule
\multirow{2}{*}{Library} &\multirow{2}{*}{Algorithm} &\multicolumn{2}{c}{{\#Unique programs}} &\multicolumn{2}{c}{{Coverage}} \\\cmidrule{3-6}
& &{Valid} &{All} &{Valid} &{All} \\\midrule
\multirow{2}{*}{\pt} &TS &\textbf{6960} &18245&\textbf{17411} &\textbf{17957} \\
&\random &6185 &\textbf{18504} &17003 &17683 \\ \midrule
\multirow{2}{*}{\tf} &TS &\textbf{5173} &\textbf{16865} &\textbf{84447} &\textbf{86536} \\
&\randombaseline &2612 &11816 &83238 &85469 \\
\bottomrule
\end{tabular}
\end{table}

\parabf{\incoder vs \codex.} Lastly, we also take a closer look at the contribution of both \codex and \incoder in generating unique test programs (\textbf{\# Unique Prog. per API}) and time cost per unique program (\textbf{Time}) in Table~\ref{tab:codex_incoder}. We observe that while \codex can provide high-quality seed programs, it is relatively slow compared to the smaller \incoder model, demonstrating the benefits of leveraging \infilling \plm{s} and evolutionary mutation to further complement the powerful but costly large \generative \plm{s} for fuzzing. %

\begin{table}[!htp]\centering
\caption{Generation efficiency of \codex and \incoder}\label{tab:codex_incoder}
\scriptsize
\begin{tabular}{l|l|rr|rrr}\toprule
\multirow{2}{*}{Library} &\multirow{2}{*}{Model} &\multicolumn{2}{c|}{\# Unique Prog. per API} &\multicolumn{2}{c}{Time per Prog. (s)} \\\cmidrule{3-6}
& &Valid &All &Valid &All \\\midrule
\multirow{2}{*}{\pt} &\codex &13.55 &23.16 &0.82 &0.48 \\
&\incoder &92.38 &450.68 &0.51 &0.10 \\ \midrule
\multirow{2}{*}{\tf} &\codex &6.85 &22.26 &1.69 &0.52 \\
&\incoder &67.17 &358.06 &0.67 &0.13 \\
\bottomrule
\end{tabular}
\end{table}

\subsection{RQ3: Detected Bugs}
Table~\ref{tab:bug } summarizes the statistics of bugs detected by \tech. In total, \tech{} detected \numTotalBugs bugs, with \numConfirmedBugs confirmed (including 19 crash and 34 wrong-computation bugs), including \numUnknownBugs confirmed as previously unknown bugs (\numFixBugs of which already fixed). Out of the \numConfirmedBugs confirmed bugs, only \numConfirmedBugsPrevWorkCanFind can be also found by the studied API-level fuzzers while none can be found by model-level fuzzers. Notably, \numConfirmedCodex confirmed bugs are found by directly using the \codex generated seeds without any mutation. 
We next present example bugs that can only be detected by \tech, as well as one rejected bug:

\begin{table}[!htp]\centering
\caption{Summary of detected bugs}\label{tab:bug }
\scriptsize
\begin{tabular}{l|r|rr|rr}\toprule
&Total &Confirmed & Unknown (Fixed) &Rejected\\\midrule
\pt &\numPtTotalBugs &\numPtConfirmedBugs &\numPtUnknownBugs (\numPtFixBugs) &\numPtRejectBugs  \\
\tf &\numTFTotalBugs &\numTFConfirmedBugs &\numTFUnknownBugs (\numTFFixBugs) &\numTFRejectBugs  \\ \midrule
Total &\numTotalBugs &\numConfirmedBugs &\numUnknownBugs (\numFixBugs) &\numRejectBugs  \\
\bottomrule
\end{tabular}
\end{table}

\begin{figure}
    \captionsetup{justification=centering}
    \centering
    \includegraphics[width=0.8\linewidth]{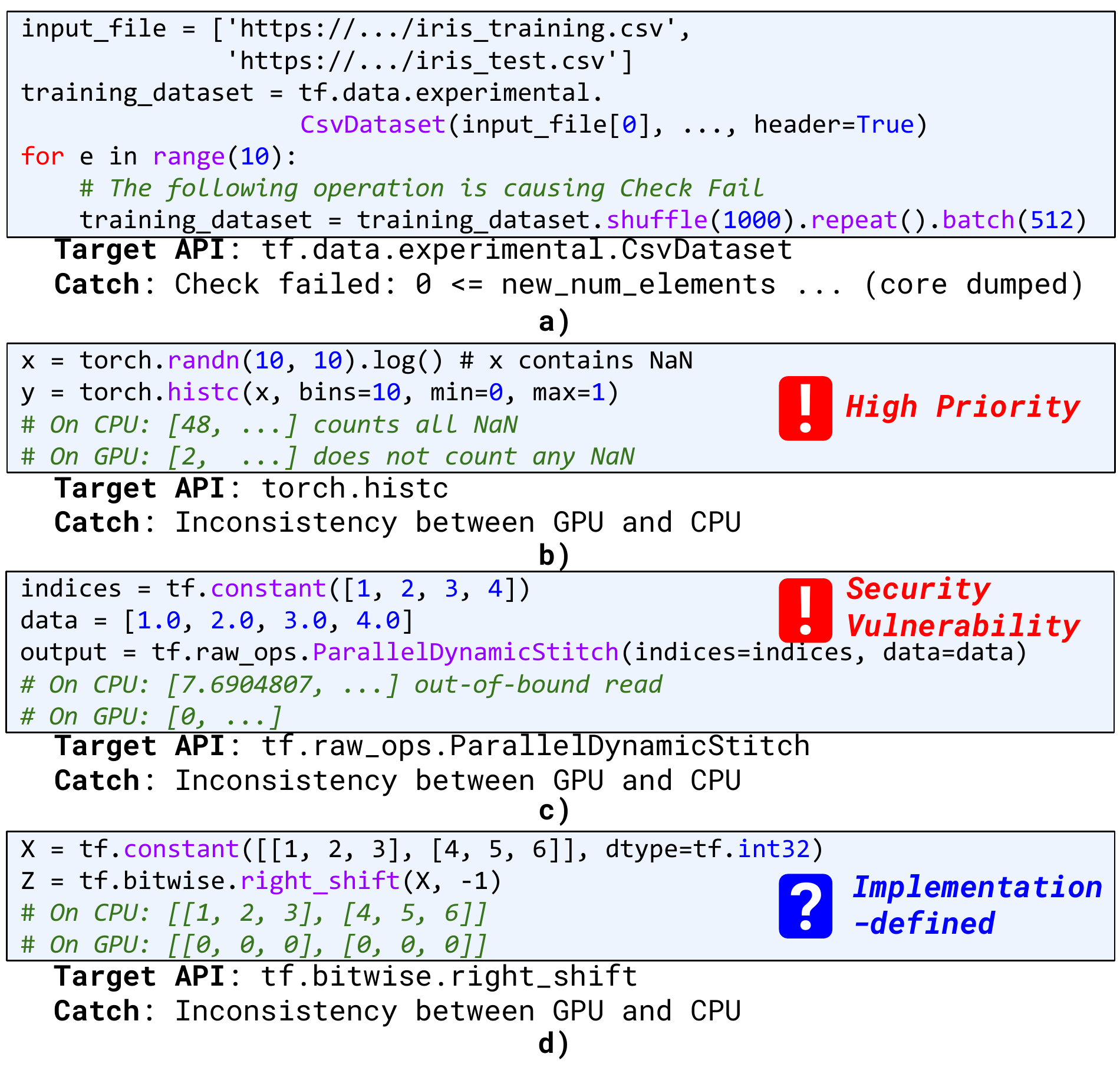}
    \caption{Bugs detected by \tech }
    \label{fig:example_detected_bugs}
\end{figure}

Figure~\ref{fig:example_detected_bugs}a shows an overflow bug found when we repeatedly batch an instance of \CodeIn{CsvDataset}. The code incorrectly crashes with \CodeIn{Check failed} message when it instead should throw a catchable overflow exception. What makes this bug hard to detect is that we first must create the dataset, which \tech{} correctly generates by calling \CodeIn{CsvDataset} (the two hyperlinks are in fact valid links to obtain data). Furthermore, the bug is triggered by using the \CodeIn{for} loop to repeatedly call the \CodeIn{batch} function. This type of unique input generation and program structure (\CodeIn{for} loop) makes it impossible for previous fuzzers to generate this test program. \tech through the use of \plm{s} can successfully generate the dataset creation code and also Python-specific code (e.g., \CodeIn{for} loop) to expose this bug.

Figure~\ref{fig:example_detected_bugs}b shows a bug in the \CodeIn{torch.histc} API where on CPU the API incorrectly counts \CodeIn{NaN} values as part of the first bin in the histogram. This bug is only found through \tech as it relies on a chained API sequence of first generating the regular random input and then applying the \CodeIn{log} function which can generate \CodeIn{NaN} values for negative inputs. Due to the silent incorrect computation, \pt developers have labeled this as a \textbf{high priority} bug.

Figure~\ref{fig:example_detected_bugs}c shows a bug when a certain output index is left unspecified for the API \CodeIn{ParallelDynamicStitch}. When running on CPU, this API can perform an out-of-bound read without throwing any exception. Although in theory, previous API-level fuzzers should be able to find this bug since it is isolated within a single API. In practice, this bug is missed since this API is not covered by any of the previous techniques. This is due to the fact that the particular API (\CodeIn{ParallelDynamicStitch}) is not commonly used. As such, previous work cannot generate valid inputs to cover this API since they rely on known valid input/API pairs obtained from databases created by scraping open-source code~\cite{freefuzz, deeprel}. \tech is able to successfully cover this API through the usage of \codex (with prompt engineering) to provide high quality seeds. Due to the potential for exploiting this silent out-of-bound read, \tf developers have labeled this bug as a \textbf{security vulnerability}.  

Since we use the differential testing oracle by comparing the values obtained when running on CPU and GPU, there could be false positive cases where inconsistencies are tolerated or intended. Figure~\ref{fig:example_detected_bugs}d shows an obvious inconsistency detected by \tech but rejected by developers. The cause is due to the inconsistency when using \CodeIn{right\_shift} with a negative value. While it is not explicitly stated in the documentation, the developers commented on the issue report that because the CPU and GPU use different lower-level shifting operators, the output result when shifting with negative values will be dependent on the implementation.

\subsection{Threats to Validity}
\mypara{Internal.} The main threat to internal validity comes from the implementation of \tech. To address this threat, the authors carefully performed testing and code review to validate that it was correctly implemented.

\mypara{External.} The main external threat to validity originates from our studied benchmarks. We mitigate this by evaluating on two most popular DL libraries: \pt and \tf. Our result shows that \tech achieves the state-of-the-art results on both libraries.%

\section{Conclusion}

We propose and implement \tech, the first approach for fuzzing DL libraries via \plmfull{s}. \tech first uses a \generative \plm (e.g., \codex) to provide high-quality seed programs through prompt engineering, and then leverages an \infilling \plm (e.g., \incoder) to mutate seed programs with an evolutionary fuzzing algorithm. Our extensive evaluation on two popular DL libraries (\pt and \tf) demonstrates that \tech significantly improves the number of covered library APIs and code coverage. Furthermore, \tech is able to detect \numTotalBugs bugs \numUnknownBugs of which are confirmed to be previously unknown. Overall, this work demonstrates a promising future of directly leveraging modern LLMs for fuzzing and testing in general.

\bibliographystyle{ACM-Reference-Format}
\bibliography{references}

\end{document}